\documentclass[prl,twocolumn,showpacs]{revtex4}
\usepackage{graphicx}
\usepackage{latexsym}
\usepackage{verbatim}%
\usepackage{amsmath}%
\usepackage{amsfonts}%
\usepackage{amssymb}

\begin{document}

\title{Current to frequency conversion in a Josephson circuit}
\date{\today }
\author{F. Nguyen}
\author{N. Boulant}
\author{G. Ithier}
\altaffiliation{Present address: Department of Physics, Royal
Holloway, University of London, Egham, Surrey, TW20 OEX, UK.}
\author{P. Bertet}
\author{H. Pothier}
\author{D. Vion}
\author{D. Esteve}
\email[Corresponding author: ]{daniel.esteve@cea.fr}
\affiliation{Quantronics Group, Service de Physique de l'\'Etat Condens\'{e} (CNRS URA
2464), DSM/DRECAM/SPEC, CEA-Saclay, 91191 Gif-sur-Yvette, France}


\begin{abstract}
The voltage oscillations which occur in an ideally current-biased
Josephson junction, were proposed to make a current standard for
metrology~\cite{likh}. We demonstrate similar oscillations in a more
complex Josephson circuit derived from the Cooper pair box: the
quantronium. When a constant current $I$ is injected in the gate
capacitor of this device, oscillations develop at the frequency
$f_{B}=I/2e$, with $e$ the electron charge. We detect these
oscillations through the sidebands induced at multiples of $f_{B}$
in the spectrum of a microwave signal reflected on the circuit, up
to currents $I$ exceeding $100~\mbox{pA} $. We discuss the potential
interest of this current to frequency conversion experiment for
metrology.

\end{abstract}
\maketitle

Exploiting the quantum properties of a current-biased Josephson
junction to make a current standard suitable for metrology was
proposed by Averin, Zorin and Likharev~\cite{likh}. This system has
a simple mechanical analog: the phase difference $\varphi$ across
the junction is equivalent to the position of a particle moving in
the Josephson potential $-E_{J}\mathrm{{cos}~\varphi}$, the voltage
across the junction to the particle velocity, and the bias current
$I$ to an applied force. The dynamics of such a particle is well
explained within the framework of the Bloch energy bands
$\epsilon_{i}(p)$ formed by the eigenstates of the particle, with
$p$ its quasi-momentum~\cite{Bloch}. It was predicted in particular
that the voltage across the junction (particle velocity) oscillates
at the Bloch frequency $f_{B}=I/2e$~\cite{likh}. These Bloch
oscillations, which provide a direct link between time and current
units, would be of fundamental interest for electrical metrology.
However, it is extremely difficult to current-bias a junction
because it requires to embed it in a circuit with a high impedance
over a wide frequency range~\cite{HavKuz,nose}. On the other hand,
it is easy to force Bloch oscillations~\cite{markus} by imposing the
quasi-momentum, that is the total bias charge $Q$ delivered to the
junction~\cite{likh}, by connecting it to a small gate capacitor
$C_{g}$ in series with a voltage source $V_{g}$, so that
$Q=C_{g}V_{g}$. This scheme cannot impose a constant current, but
 can deliver alternatively two opposite values of the current
$\mbox{d} Q/\mbox{d} t=\pm I$. In this Letter, we report experiments
using this procedure and demonstrating oscillations at the Bloch
frequency $f_{B}=I/2e$ in a Josephson circuit that allows their
detection. Our setup, shown in Fig.~1a, is based on a modified
Cooper pair box~\cite{cpb}, the quantronium~\cite{VionSc}. We show
how this new current-to-frequency conversion method exploits the
quantum properties of the circuit. We also discuss its interest in
metrology of electrical currents, for which electron
pumping~\cite{pump,pekola} and electron counting~\cite{per} have
also been proposed.

\begin{figure}[ptbh]
\begin{center}
\includegraphics[width=0.9\columnwidth,clip]{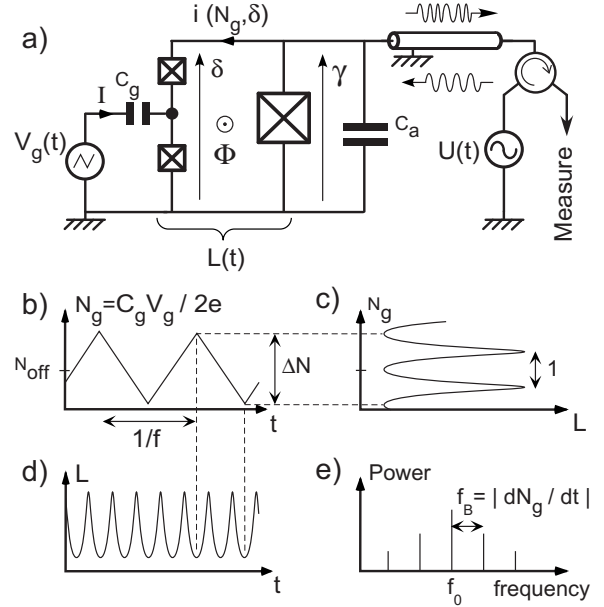}
\end{center}
\caption{Operating principle of the quantronium circuit for the
production and detection of Bloch-like oscillations. The circuit(a)
is a split Cooper pair box with a probe junction for the detection
of the oscillations that develop when the gate charge is swept
linearly. When the linear sweep is replaced by a triangular sweep
(b) with extrema corresponding to symmetry points of the inductance
modulation pattern (c), the time variations of the inductance (d)
are the same as for a continuously increasing linear sweep. This
modulation manifests itself as sidebands in the spectrum (e) of a
microwave signal
reflected onto the circuit.}%
\label{fig1}%
\end{figure}

The quantronium device~\cite{VionSc,thesis} is a split Cooper pair
box that forms a loop including also a probe junction. The box
island with total capacitance $C$ is defined by two small junctions
having Josephson energies $E_{J}(1+ d)/2$ and
 $E_{J}(1- d)/2$,  $d$ being an asymmetry coefficient.
  The superconducting phase $\widehat{\theta}$ of this island,
conjugated to the number $\widehat{N}$ of extra Cooper pairs inside,
forms the single degree of freedom of the box~\cite{cercle}. The
third larger junction with critical current $I_{0}$, in parallel
with an added on-chip capacitor $C_{a}$, forms a resonator with
plasma frequency $f_{p}$ in the $1-2~\mbox{GHz} $
range~\cite{zorin}. Since this frequency is always smaller than the
box transition frequency, we treat the phase difference ${\gamma}$
across the probe junction as a classical variable.  The same thus
holds for the phase difference $\delta =\gamma+\phi/\varphi_{0}$
across the two box junctions in series, with ${\phi}$ the magnetic
flux applied through the loop, and $\varphi_{0}=\hbar/2e$. The
control parameters of the split-box are $\delta$ and the reduced
gate charge $N_{g}=C_{g}V_{g}/2e$, with $C_{g}$ the island gate
capacitance, and $V_{g}$ the gate voltage. The Hamiltonian of the
box writes
\begin{equation}
\widehat{H}=E_{C}(\widehat{\mathbf{N}}-N_{g})^{2}-E_{J}\cos\frac{\delta}%
{2}\cos\widehat{\mathbf{\theta}}+dE_{J}\sin\frac{\delta}{2}\sin
\widehat{\mathbf{\theta}}~\text{,}\label{splitBoxHam}%
\end{equation}
with $E_{C}=(2e)^{2}/2C$.

The eigenenergies $\epsilon_{i}(N_{g},\delta)$ vary periodically
with $N_{g}\>(\mbox{period 1} )\>$ and $\delta\>$(\mbox{period}
$2\pi $)~\cite{thesis}. The experiment consists in imposing a linear
variation of the reduced quasi-momentum $N_{g}$ that induces a
periodic evolution of the quantum state along the first Bloch band
$\epsilon_{0}(N_{g},\delta)$, at the Bloch frequency $f_{B}=\mbox{d}
N_{g}/\mbox{d} t=I/2e$. Therefore, the
current $i(N_{g},\delta)=\varphi_{0}^{-1}\partial\epsilon_{0}(N_{g}%
,\delta)/\partial\delta$ through the two small junctions, the
associated effective inductance for small phase excursions
\begin{equation}
L(N_{g},\delta)=\varphi_{0}^{2}\left(\frac{\partial^{2}\epsilon
_{0}(N_{g},\delta)}{\partial\delta^{2}}\right)^{-1},\label{ind}%
\end{equation}
and hence the admittance $Y(\omega)=j\left[
C\omega-I_{0}/\varphi_{0}\omega-1/L(N_{g},\delta)\omega\right]$ as
seen from the measuring line, vary periodically. We measure this
admittance $Y(\omega)$ by microwave reflectometry, as for the
RF-SET~\cite{rfset}. In our experiment, we apply a triangular
modulation of the gate signal centered on
$N_{g}=N_{\mathrm{{off}}}$, with peak to peak amplitude $\Delta N$,
and with frequency $f_{g}$ corresponding to $\mbox{d} N_{g}/\mbox{d}
t=\pm I/2e$. Due to the symmetry properties of the quantum states
with respect to $N_{g}$, the inductance \label{induct} varies as for
a linear sweep, as shown in Fig.~1, provided that the extremal
values of $N_{g}$ are integer or half-integer, with a Bloch
frequency $f_{B}=2\Delta Nf_{g}$. In order to obtain the largest
gate charge modulation of the inductance, the phase is adjusted at
$\delta\approx\pi$ with the flux ${\phi}$. When a small microwave
signal at frequency $f_{0}$ is sent on the measuring line, the
periodic modulation of the reflection factor yields sidebands in the
spectrum of the reflected signal~\cite{rfset}, shifted from the
carrier by multiples of $f_{B}$, and called Bloch lines~\cite{ms+s}.
Due to the periodic excitation, the stationary outgoing amplitude
can be written as a series:
\begin{equation}
v_{out}(t)=\sum_{k}v_{k}\>\mbox{exp} [2i\pi(f_{0}+kf_{g})t].\label{out}%
\end{equation}
The circuit equations and the loop-current expression ~\cite{VionSc,thesis}
allow to calculate all sideband amplitudes $v_{k}$, which get smaller and
become asymmetric $(v_{-k}\neq{v_{k}})$ when the sideband frequencies depart
from the resonance.

\begin{figure}[ptbh]
\begin{center}
\includegraphics[width=0.95\columnwidth]{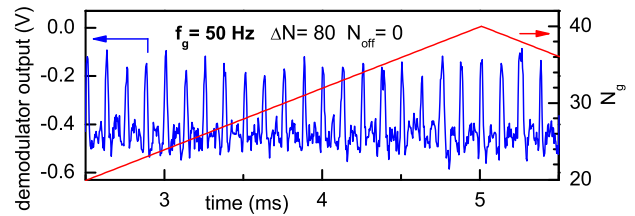}
\end{center}
\caption{(Color online) Demodulated output signal (left scale)
recorded with a 300~Hz~-~30~kHz bandwidth during a $3~\mathrm{{ms}}$
time window, when a triangular wave voltage corresponding to a Bloch
frequency $f_{B}=8~\mbox{kHz} $ is applied to the gate (right
scale). Each period corresponds to the
injection of one extra Cooper pair.}%
\label{fig2}%
\end{figure}

The sample was fabricated using electron-beam lithography
~\cite{VionSc,thesis} and aluminum deposition and oxidation. In
order to avoid quasiparticle poisoning, the island was made thinner
than the leads (13~nm and 42~nm respectively), and gold
quasiparticle traps were used~\cite{sset}. In the present
experiment, a sizeable asymmetry $d$ was introduced on purpose in
order to maintain a large gap $G_{0}=\epsilon_{1}-\epsilon_{0}$ at
$(N_{g}=1/2,\delta=\pi)$, which avoids microwave driven transitions
towards excited bands. The sample was placed in a sample-holder
fitted with microwave transmission lines, at a temperature
$T\simeq30~\mbox{mK} .$ The gate was connected to a $250~\mbox{MHz}
$-bandwidth RF line. The microwave signal, after reflecting on the
probe junction, went through 3 circulators before being amplified by
a cryogenic amplifier with noise temperature $T_{N}=2.2~\mbox{K} $,
and a room temperature amplifier. The signal was then either
demodulated with the CW input signal, or sent to a spectrum
analyzer. In the latter case, the applied power was
$\approx-132~\mbox{dBm} $, corresponding to phase excursions smaller
than $\pm0.1~\mathrm{{rad}}$, and the total measurement gain was
$\approx 88~\mbox{dB} $. As a function of $N_{g}$ and $\delta$, the
plasma resonance varied in the range $1.11-1.21~\mbox{GHz} $,
slightly below the circulator bandwidth, which yielded an extra
attenuation of the signal due to a spurious interference with the
leakage signal through circulator~\cite{atten}. Fitting the
variations of the resonance yielded $f_{p}=1.19~\mbox{GHz} $,
 $Q\approx17$, $E_{C}=1.42\pm{0.2}~\mathrm{k_{B}K}$,
 $E_{J}=2.88\pm{0.2}~\mathrm{k_{B}K,}$ and
$d=0.15\pm{0.03}$ leading to $G_{0}\approx~7hf_{0}$.

\begin{figure}[ptbh]
\begin{center}
\includegraphics[width=0.95\columnwidth,clip]{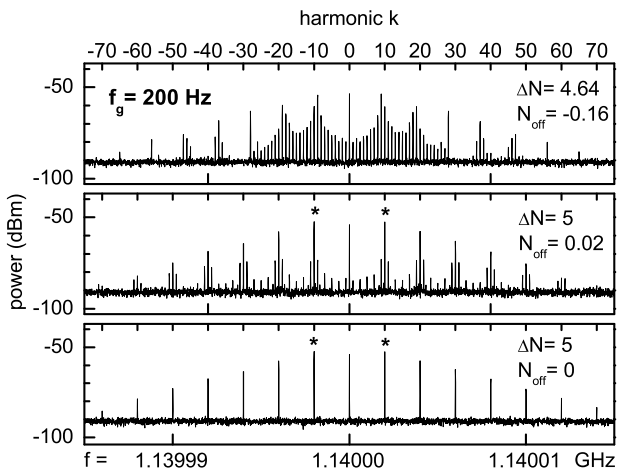}
\end{center}
\caption{Spectrum of a reflected CW signal at $1.14~\mbox{GHz} $ for
different triangular wave gate modulation patterns, at frequency
$f_{g}=200~\mbox{Hz} $. The spectrum consists of sidebands shifted
by $kf_{g}$ from the carrier. Progressive tuning of the amplitude
$\Delta N$ and of the offset $N_{\mathrm{{off}}}$ yields a spectrum
consisting only of Bloch lines shifted from the carrier by a
multiple of $f_{B}=~2\Delta N f_{g}$. The Bloch lines of
order 1 ($k=\pm2\Delta N$) are marked by an asterisk.}%
\label{fig3}%
\end{figure}

The reflected signal demodulated with the carrier is shown in Fig.~2
for a triangular gate voltage corresponding to $f_{B}=8~\mbox{kHz}
$. Due to noise, such time-domain measurements could only be
performed within a 100~kHz bandwidth~\cite{bandwidth}. In the
following, the reflectometry spectra are taken with a
$1~\textrm{Hz}$ bandwidth resolution.

A series of spectra recorded at $f_{0}=1.14~\mbox{GHz} $ with $f_{g}%
=200~\mbox{Hz} $, and taken with progressively tuned gate sweep signal
amplitude and offset is shown in Fig.~3: when $\Delta N$ and $N_{\mathrm{{off}%
}}$ are tuned as sketched in Fig.~1, the spectrum consists only of
Bloch lines, as predicted, with linewidth limited by the spectrum
analyzer.

An example of comparison between the measured and predicted sideband
amplitudes when $N_{\mathrm{{off}}}$ or $\Delta N$ is varied is
shown in Fig.~4 for $f_{g}=1~\mbox{kHz} $. The measured amplitudes
are well accounted for by the solution of Eq.~\ref{ind}, but for the
carrier, which suffers from the spurious interference effect already
mentioned~\cite{atten}. Although the overall agreement for the
sidebands, and in particular the cancelation of some of them at
particular offsets and amplitudes, demonstrate the phase-coherence
of the measured signal, it does not prove that the quantronium
undergoes a perfect coherent adiabatic evolution of its ground-state
while $N_{g}$ and $\delta$ are varied: incoherent
excitation/de-excitation processes are for instance not excluded.
Because of the opposed inductance modulation in the excited state,
they would only reduce the amplitude of the Bloch lines in
proportion of the time spent in this state.

\begin{figure}[ptbh]
\begin{center}
\includegraphics[width=0.95\columnwidth,clip]{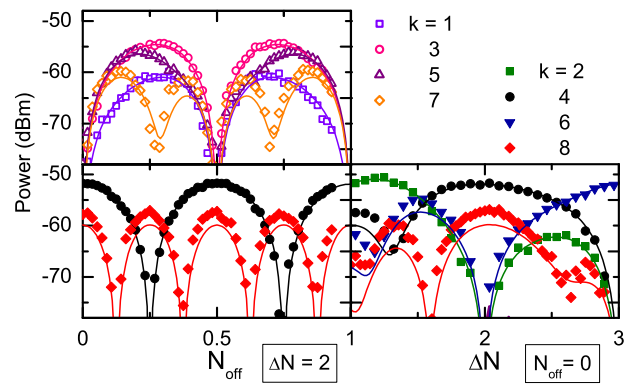}
\end{center}
\caption{{(Color online) Comparison of the measured (symbols) and
calculated (lines) sideband amplitudes as a function of the sweep
offset $N_{\mathrm{{off}}}$ and amplitude $\Delta N$, for
$f_{g}=1~\mbox{kHz} $}. Left panels: Offset dependence for $\Delta
N=2$; the observed sidebands correspond as predicted to odd
multiples of $f_{g}$ (top) and to the Bloch line $k=4$ and its
harmonics $k=4n$, with period $1/(2n)$ in $N_{\mathrm{{off}}}$
(bottom). Right panel: amplitude dependence for
$N_{\mathrm{{off}}}=0$. The Bloch line corresponds to harmonic 4 at
$\Delta N=2$, and to harmonic 6 at $\Delta N=3.$ Calculated curves
were shifted by $45~\mbox{dB} $ (estimated value was $44~\mbox{dB}
$) to best match the experimental Bloch line of order
1.}%
\label{fig4}%
\end{figure}

Spectra obtained at larger frequencies $f_{g}$ and larger amplitudes
$\Delta N$, corresponding respectively to currents and Bloch
frequencies $I=32~\mbox{pA} $, $f_{B}=100~\mbox{MHz} $, and
$I=130~\mbox{pA} $, $f_{B}=408~\mbox{MHz} ,$ are shown in Fig.~5.
These results demonstrate that Bloch oscillations persist at Bloch
frequencies larger than the resonator bandwidth
$f_{p}/Q\approx{70~\mbox{MHz} }$, even though Bloch lines become
weaker. The successful current to frequency conversion performed up
to currents $I>100~\mbox{pA} $ using Bloch oscillations is the main
result of this work. However, the amplitudes of the Bloch lines at
these high frequencies are smaller than predicted by the model, and
additional sidebands are present, which we attribute to the rounding
of the gate triangular wave signal at its turning points, and to
drifts of the gate-charge due to background charge noise. In
principle, the theoretical maximum current is limited by two
fundamental phenomena. First, the conversion mechanism requires
$f_B\ll f_0$, with $f_0\ll G_0/h$ to avoid multi-photon excitations.
Second, the current must be low enough to avoid Zener transitions at
$N_g=1/2$ where the gap is minimum. For the present experiment, the
Zener probability $p_Z=\exp(-\pi^{2}{G_0}^{2}/hsf_B)\approx\exp(-13~
\textrm{GHz}/f_B)$, with $s$ the slope of
$\epsilon_{1}(N_{g})-\epsilon_{0}(N_{g})$ away from $N_{g}=1/2$, is
negligible.

\begin{figure}[ptbh]
\begin{center}
\includegraphics[width=0.95\columnwidth,clip]{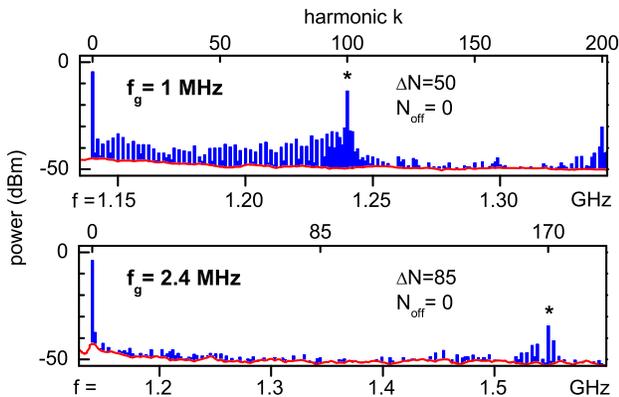}
\end{center}
\caption{ (Color online) Amplitude of the sidebands at positive harmonics (top
scale) of the gate frequency $f_{g}$. Top panel: $\Delta N=50$, $f_{g}%
=1~\mbox{MHz} $ ($I=32~\mbox{pA} $,$f_{B}=100~\mbox{MHz} $) ; bottom
panel:
$\Delta N=85$, $f_{g}=2.4~\mbox{MHz} $ ($I=130~\mbox{pA} $, $f_{B}%
=408~\mbox{MHz} $). The Bloch line of order 1 is marked by an asterisk. The
continuous line is the noise level. We attribute the presence of non Bloch
lines to the imperfections in the gate signal, and to charge noise.}%
\label{fig5}%
\end{figure}

An important application of these experimental results could be to
establish a direct link between a DC current and a frequency through
the Bloch frequency $f_{B}=I/2e,$ in order to close the triangle of
quantum metrology~\cite{tri}. Such an experiment would aim at
measuring the current $I_{H}$ passing through a Quantum Hall bar
device in terms of a rate $\dot{N_{H}}$ of transferred Cooper pairs,
in order to check the consistency of the Quantum Hall effect (QHE)
with the AC Josephson effect. Using this latter effect, the Hall
voltage $V_{H}=(h/e^{2})I_{H}$ across a QHE bar can indeed be
related to a frequency $f_{H}$ through the relation
$(h/e^{2})I_{H}=(h/2e)f_{H}$. If the description of both QHE and
Josephson experiments is exact, one predicts $\dot{N_{H}}=f_{H}/4$.
A consistency check of this relation at the $10^{-8}$ level is
presently a major goal in metrology because, in conjunction with a
metrological realization of the mass unit by a Watt-balance
experiment~\cite{bal}, it would provide a serious basis for a
redetermination of the SI unit system in terms of electrical
experiments involving only fundamental constants. The large current
$I_{H}\approx1~\mu\mbox{A} $, needed for QHE experiments, can be
transposed to a smaller range $0.1-1~\mbox{nA} $ using topologically
defined transformers~\cite{cccc}. This current range, which is still
beyond reach of single electron pumps~\cite{pump} or of direct
electron counting experiments~\cite{per}, can be accessed with the
sluice Cooper pair pump~\cite{pekola}, with Bloch oscillations in a
single Josephson junction~\cite{HavKuz}, or with the method
demonstrated here provided it can be used with a true DC current. In
the last two cases, the impedance of the current source as seen from
the single junction or from the box needs to be larger than
$R_{Q}=h/4e^{2}$ to preserve single Cooper pair effects, and
temporal fluctuations have to be small enough to obtain
 narrow Bloch lines  enabling an accurate
 measurement of their frequency. Using for instance a resistive
 bias yields a Bloch linewidth of the order of
 the Bloch frequency~\cite{HavKuz}, due to thermal
 fluctuations of the self-heated bias resistor.
Developing a suitable current source for charge injection is thus a
challenging prerequisite to metrology experiments based on Bloch
physics. High impedance dissipative linear Josephson arrays have
already been used to demonstrate indirectly Bloch
oscillations~\cite{nose}. Combining ohmic, inductive, and Josephson
elements, and possibly non-equilibrium cooling
techniques~\cite{cool}, might provide an adequate low-noise high
impedance.

In conclusion, we have demonstrated the conversion of a current $\pm
I$ to a frequency $f_{B}=I/2e$ in a Josephson device biased through
a small capacitor, through the production of ultra narrow sidebands
in the spectrum of a reflected microwave signal. This new method,
which reaches a current range $I>0.1 \mbox{nA} $, would be extremely
appealing for metrology if operated with a DC current.

\vskip 0.3cm
We acknowledge discussions with M. Devoret, F.
Piquemal, W. Poirier, N. Feltin, and within the Quantronics group.
This work has been supported by the European project Eurosqip and by
the C'nano grant 'signaux rapides'.



\begin{thebibliography}{9}
\bibitem {cercle}In contrast to the case of a current-biased junction, the
phase $\widehat{\theta}$ is defined on a circle due to the quantization of
$\widehat{N}$.
\bibitem {atten}The circulator leakage requested to explain our data varied
from $-7~\mbox{dB} $ to $-23~\mbox{dB} $ over the plasma frequency excursion
range, which is compatible with room temperature measurements.
\bibitem {bandwidth}Note that a larger bandwidth was recently obtained in
related electrometry setups~\cite{sset}.
\end{thebibliography}
\end{document}